\documentclass[aps,prl,amsmath,twocolumn]{revtex4}
\usepackage{bm}
\usepackage{graphicx}
 \usepackage{amssymb}

\def \d{\partial}

\def \bu{{\bf u}}

\def \bxi{\boldsymbol{\xi}}

\def \br{{\bf r}}
\def \bA{{\bf A}}
\def \bQ{{\bf Q}}
\def \be{{\bf e}}
\def \bx{{\bf x}}
\def \bl{{\bf  l}}

\def \bR{{\bf R}}
\def \bD{{\bf D}}
\def \bZ{{\bf Z}}

\begin{document}

\title{Material surfaces in stochastic flows: integrals of motion and intermittency }

\author{ A.S. Il'yn$^{1,2}$, A.V. Kopyev$^*$, V.A. Sirota, K.P. Zybin$^{1,2}$
\thanks{Electronic addresses:  asil72@mail.ru, kopyev@lpi.ru, sirota@lpi.ru, zybin@lpi.ru}}
\affiliation{$^1$ P.N.Lebedev Physical Institute of RAS, 119991, Leninskij pr.53, Moscow,
Russia \\
$^2$ National Research University Higher School of Economics, 101000, Myasnitskaya 20, Moscow,
Russia}

\begin{abstract}
We consider the  line, surface and volume elements of fluid in stationary isotropic incompressible
stochastic flow in  $d$-dimensional space and investigate the long-time evolution of their
statistic properties.
 We report the discovery
of a family of  $d!-1$  stochastical integrals of motion that are universal in the sense their
explicit form does not depend on the statistics of velocity. Only one of them has been discussed
previously.
\end{abstract}

\maketitle

\section{I. Introduction}

The evolution of material lines and surfaces in a turbulent flow is
important for the theory of turbulence
and 
turbulent transport  \cite{2inPope2, Pope2, FGV, MY75}. It
 provides inherently geometric view on turbulent mixing \cite{Dimotekis, Bentkamp}. The
study of material elements is of intrinsic interest and practical value for many applications;
e.g., the evolution of infinitesimal material lines is identical to that of frozen magnetic field
in highly conducting media \cite{Chertkov, KKISZ}, and the material surfaces trace the
constant-property surfaces of passive scalars in the limit of negligible molecular diffusivity
\cite{BalkFoux, Celani, ISZscal} or flamelet propagation for flame speeds slow
compared to the Kolmogorov scale \cite{Sab}, or salinity waves in oceans \cite{2inBent}. So, good understanding of material elements
evolution is also necessary for problems of turbulent dynamo and combustion.

On the other hand, stochastical integrals of motion are one of the most important instruments to
investigate systems far from equilibrium, as turbulent flow is.
They help  to reveal the basic mechanisms of turbulence \cite{FGV}.
In this article, we find new universal (i.e., independent of velocity statistics) integrals of
motion for material elements.

The evolution of material line and area elements has been analyzed by many authors
\cite{Batchelor52, Kraichnan74, Bentkamp, 2inPope2, Pope2, Pope1, Celani}
theoretically, experimentally and numerically, under different assumptions on the velocity field.
Mathematically, infinitesimal material elements correspond to differential forms. Physically,
infinitesimality corresponds to scales much less than the Kolmogorov viscous scale length
(so-called Batchelor regime). At these scales,  separation of trajectories is exponential, and same
is the evolution of material line element length. The velocity field at such scales is linear, it
is determined by the velocity gradient tensor $A_{ij}(t)=\d_j v_i$. So, all statistical properties
of the differential forms (and, hence, of lengths, squares etc.) are completely determined by the
statistics of $A_{ij}(t)$ along a liquid particle trajectory.

There exists an infinite set of time-invariant configurations; their explicit form generally
depends on the statistics of the velocity gradient tensor along particle trajectories. However, in
\cite{Furst, zeld-integral} there was found a non-trivial integral of motion that is universal: its
expression does not depend on details of  velocity statistics. It appears that there exists a
family of
such integrals,  all of them are  averages (or hypersurface integrals) of some powers of
 the differential forms' absolute values. In this paper we find them all; for $d$ dimensional flow,
 there are $d!-1$ invariants.

 The developed technics also allows us to find various non-trivial invariants
 expressed by ratios of different averages.

The existence of these universal invariants is essentially nontrivial. It is a consequence of
statistical
 isotropy of the flow, in combination with very particular properties of the evolution operator of
material elements.

In the next Section we formulate the problem statement and the main results. In Sections III, IV
we proceed to accurate analysis of the $d$-dimensional case. In Section~V, we derive the stochastic
integrals of motion. In the last Section, we discuss briefly some
 properties and manifestations of the discovered invariants, and some other possibilities to find
 integrals of motion.

 \section{II. Problem statement and results}

 Consider $d$-dimensional space filled with fluid (continuous set of particles) that flows
 according to the equation
\begin{equation}  \label{1}
\frac{d \br (t)}{dt} = \bu (t, \br (t))
\end{equation}
where $\bu$ is some random stochastically isotropic and homogenous (hereafter isotropic) stationary
vector field with finite ($<\infty$) correlation time and length. Its statistics is assumed to be
known. For instance, $\bu$ may obey the Navier-Stokes equation with random forcing \cite{Novikov},
or one can use the Gaussian delta-correlated velocity field (Kraichnan model
\cite{Kraichnan}). We also assume the incompressibility condition  $\nabla \cdot \bu =
0$.

We are interested in the evolution of material lines and ($k < d$ -dimensional) hypersurfaces. So,
we introduce a coordinate grid that is orthogonal at the initial moment,
$$
\br (0,\bx) = \sum \be_{i} x^i \ , \qquad \be_{i}\cdot \be_{j} = \delta_{ij}
$$
Here $\{ \be_i \}$ is a set of orthonormal orts, and ${\bf x} = \{ x^i \}$ is the Lagrangian marker
of each particle. The grid is trapped in the turbulent flow; the position of every point of the
grid changes according to Eq. (\ref{1}), and the coordinate lines and planes become deformed and
bent.

Then, time evolution of the tangent vectors
\begin{equation}  \label{ldef}
\bl_i (t,\bx) = \frac{\d \br (t,\bx) }{ \d x^i}
\end{equation}
 is described by the equation
\begin{equation}  \label{2}
\frac{d \bl_i}{dt} = \frac{d}{dt} \frac{\d \br}{\d x^i} = \frac{\d}{\d x^i} \bu = {\bf A} \bl_i \ ,
\quad \bl_i(0) = \be_i
\end{equation}
where ${\bf A}(t,\br)$ is the velocity gradient tensor, $A_{ij}=\d u_i / \d r_j$; the time
derivative is taken along the particle trajectory, i.e., at some constant $\bx$.

The tangent vector field $\bl_i (t,\bx)$  is also called Cartan 1-form and describes the evolution
of an infinitesimal length element: the length of a segment of the frozen Lagrangian coordinate
line $x^{j\ne i}=const$,  $x^i \in L_0$ is
$$
L= \int \limits _{L_0} \| \bl_i (t,\bx) \| dx^i  \qquad (no \ summation)
$$
The square of a segment of frozen Lagrangian coordinate plane is
$$
S = \int \limits_{\sigma_0} \| \bl_i \wedge \bl_j \| dx^i dx^j \qquad (no \ summation)
$$
So, the surface element is described by the Cartan 2-form $s^{(2)}_{ij} (t,\bx) = \bl_i \wedge
\bl_j $, where $\wedge$ denotes the outside (vector) product.

Generally, the evolution of $k$-dimensional  Lagrangian coordinate hypersurface is described by the
Cartan $k$-form
\begin{equation}  \label{Cartanform}
S^{(k)}_{i_1..i_k} (t,\bx) = \bl_{i_1} \wedge ... \wedge \bl_{i_k}
\end{equation}

 From isotropy it follows that all Lagrangian coordinate planes are equivalent: averages of all quantities
do not depend on their orientation and position. So, one can restrict the consideration to the set
$$
\bl_1,  \    \bl_1 \wedge \bl_2 , \  ..., \  \bl_1 \wedge \bl_2 \wedge ... \wedge \bl_d
$$
and investigate the time evolution of the norms
\begin{equation}  \label{norms}
s_1 = \| \bl_1 \| , \  s_2= \| \bl_1 \wedge \bl_2 \|, \  ..., \  s_{d-1} = \| \bl_1 \wedge \bl_2
\wedge ... \wedge \bl_{d-1} \|
\end{equation}
For incompressible flow, $s_d$ is constant.

We require the velocity field to satisfy the following condition:
 the statistics of $\bA(t,\bx) = \bA (t, \br (t,\bx))$ taken along a $\bx$-particle trajectory is
 stationary with finite correlation time, and same for all trajectories.
 For incompressible flow, this condition holds as a result of isotropy.

 Now we can formulate the main result of the paper.

Let $i \rightarrow \pi(i), i=1..d $ be a permutation ($\forall 1\le i \le d : 1 \le \pi(i) \le d, \
\pi(i) \ne \pi(j)$ if $i\ne j$). Then in long-time asymptotics, there exists a
 stochastic integral of motion:
\begin{equation} \label{result}
\left \langle s_1 ^{\pi(2)-\pi(1)-1} s_2 ^{\pi(3)-\pi(2)-1}...s_{d-1} ^{\pi(d)-\pi(d-1)-1}
 \right \rangle  = const
\end{equation}
to logarithmic accuracy.

 The average in (\ref{result}) can be taken either over the ensemble of
realizations of $\bu (\br)$ for some chosen point of the Lagrangian grid, or over any Lagrangian
coordinate hyperplane in a given realization.

There are $d!-1$ non-trivial permutations, thus we get the same number of universal stochastic
conservation laws. We stress that the only essential restrictions we use are the ones listed below
Eq.(\ref{1}); details of statistics do not matter.

In particular, for cyclic permutations we get $d-1$ integrals of motion:
\begin{equation} \label{cyclic-inc}
\left \langle s_k^{-d} \right \rangle = const  \qquad  k = 1 \dots d-1
\end{equation}
  First of these integrals of motion is
well known: it was found in \cite{zeld-integral}.

With account of isotropy, the ensemble averages can be written as integrals of some powers of
$k$-dimensional hypersurface density $\sigma_k=s_k^{-1}$ over the $k$-hypersurface moving along
with the flow:
\begin{equation} \label{densities}
\int \sigma_k ^{d+1} dS = const
\end{equation}

\subsection{Note on intermittency}
As we will show below,  in incompressible flow the length of material line increases exponentially
on average, as well as square of material surface etc. To the contrary,  $\left \langle s_1
^{-d}\right \rangle = const$,  $\left \langle s_2 ^{-d}\right \rangle = const$,...  For negative
degrees $0>a>-d$, $\left \langle s_k^a \right \rangle$ decreases exponentially as a function of
time, while for $a<-d$ $\left \langle s_k^a \right \rangle$ increases. This illustrates
intermittency of the turbulent flow: averages over positive degrees of $s_k$ are mainly contributed
by the regions where material elements  stretch most intensively, while averages over negative
degrees are dominated by even more rare regions where the material elements undergo exponential
contraction. The integrals of motion correspond to a balance between high speed of contraction and
low probability (frequency) of fast-contracting elements.

\section{III. Cartan forms in $d$ dimensions}
Consider the evolution matrix $\bQ(t)$: $\br(t,\bx) = \bQ \br (0,\bx)$. Then from (\ref{1}) it
follows that
$$
d \bQ/dt = \bA(t) \bQ(t) \ , \quad \bQ (0) = {\bf I}
$$
The formal solution to this equation can be written by means of T-exponent,
$$
\bQ(t) =  \mathcal{T} \left\{ e^{\int \limits_0^t \bA(t')dt'} \right\}
$$
However, the matrices
$\bA(t')$ taken at different time moments do not commutate, which causes immense difficulties and
makes the explicit expression for $\bQ(t)$ impossible. To deal with this stochastic matrix
equation, it is convenient to consider the Iwasawa decomposition of the matrix $\bQ$:
$$
\bQ (t) = {\bf R} {\bf D} {\bf Z} \ ,
$$
where $\bR$ is orthogonal matrix, $\bD$ is positive diagonal, and $\bZ$ is upper-triangular
unipotent matrix:
$$
{\bf R R}^T = \hat{\bf I} \ , \ \ D_{ij} = \delta_{ij} D_i , \ D_i>0 , \ \  Z_{i>j}=0 \ , Z_{jj}=1
$$
From the multiplicative  Oseledets theorem \cite{Oseledets} it follows that almost surely, there
exist the limits
$$  
\mathop {\lim }\limits_{t \to \infty } \left( \frac 1t  \ln D_k \right)  =  \lambda_k \ , \qquad
\lambda_1 \ge \dots \ge \lambda _d
$$ 
Now, the Cartan forms (\ref{Cartanform}) can
be written as
$$
S^{(k)}_{1..k} (t,\bx)= \bQ \be_{1}  \wedge \dots \wedge \bQ \be_{k} ={\bf R} D_1 \be_{1} \wedge
\dots \wedge {\bf R} D_k \be_{k}
$$
Here we make use of the fact that $\bZ$ is upper-triangular.  So the norms (\ref{norms}) take the
form
\begin{equation} \label{s-D}
s_k = D_1 D_2 \dots D_k
\end{equation}
So, for our purposes we are interested only in the evolution of the $D_i$ components.

\section{IV. Generalized Lyapunov exponents}

To calculate the correlators of $s_k$ we need the averages like $\langle D_1^{m_1}\dots
D_d^{m_d}\rangle$. One can show \cite{PRE22} that each $D_i$ satisfies the equation
\begin{equation} \label{D-xi}
d { D_i} /dt = { \xi_i D_i} ,
\end{equation}
where  $\bxi (t) = \{ \xi_1,\dots,\xi_d \} $ is a set of stationary random processes that depend on
$\bA(t)$ in a rather complicated way.$^4$\footnotetext[4]{More precisely, $\bxi$ is the diagonal part of the statistically stationary
random matrix ${\bf R}^{-1}\bA \bR$.}
For each component  $D_i$, the solution of Eq. (\ref{D-xi})~is
$$
D_i(t) = e^{\int \limits_0^t \xi_i(t')dt'} = e^{t \bar{\xi}_i(t)} \ ,
$$
where $\bar{\bxi}(t)= \frac 1t \int_0^t \bxi(t')dt'$ is the time average of $\bxi$.

We assume that $\bxi(t)$ satisfies the large deviations principle
 \cite{Varadhan}, i.e., that the joint
probability density of all $\bar{\xi}_i(t)$  at large $t\to \infty$ satisfies the relation:
$$
\mathcal{P}_{\bar{\bxi} } (a_1,\dots,a_d) \equiv \left \langle \prod \limits_i
\delta(\bar{\xi}_i(t) - a_i)\right \rangle \sim e^{-t J(a_1,\dots,a_d) }
$$
Here the angle brackets denote the ensemble average, and the sign
 $\sim$ means that
$$
\lim \limits_{t \to \infty} \frac 1t  \ln {\cal P}_{\bar{\bxi} } (a_1,\dots,a_d)  = -
J(a_1,\dots,a_d)
$$
The function $J$ is called Cramer function, or effective action. It is concave, and has the minimum
$J_{min}=0$ at ${\bf a}_{min}=\langle \bxi \rangle$. Then
$$
\langle D_1^{m_1}\dots D_d^{m_d} \rangle = \left \langle e^{\sum m_i  \int\limits_0^t {\xi}_i(t')
dt'}\right \rangle \sim \int e^{t(\sum m_i a_i -J({\bf a}))} d{\bf a}
$$
and there exists the limit
\begin{equation} \label{12}
\lim \limits_{t\to \infty}  \frac 1t \ln \left \langle D_1^{m_1}\dots D_d^{m_d} \right \rangle =
w_{\xi}(m_1,\dots, m_d) \ ,  \quad m_k\in \mathbb{R} \  ;
\end{equation}
where $w_{\xi}$ is the Legendre transform of $J$. The function $w_{\xi}$ is called generalized
Lyapunov exponent (GLE) \cite{CrisantiPaladinVulp,PRE22}.

Generally, the statistics of $\bxi$ is not determined by the statistics of $\bA$ at the same moment
of time: it also depends on the prehistory. However, in \cite{JOSS1,PRE22} it was shown that  in
the case of statistically isotropic $\bA (t)$ there exist a simple relation between the statistics
of $\bxi$ and $\bA$, namely,
\begin{equation} \label{main}
 w_{\xi} (m_1,\dots, m_d) =
w_A(m_1+\eta_1, \dots, m_d+\eta_d) - w_A(\eta_1, \dots, \eta_d)
\end{equation}
where $w_A$ is the cumulant-generating function corresponding to the diagonal elements of the
matrix $\bA$,
$$
w_A (m_1,\dots m_d)=\lim \limits_{t\to \infty} \frac 1t \ln \left \langle e^{ \int (m_1
A_{11}+\dots + m_d A_{dd}) dt } \right \rangle \ ,
$$
and $\eta_k$ is a set of constants defined by
\begin{equation} \label{eta}
\eta_k = (d+1)/2 - k
\end{equation}
 This relation allows to calculate the statistical characteristics  of $\bD$ for given statistics of $\bA$;
   in particular, to find all statistical moments.

\section{V. Stochastic integrals of motion}

From (\ref{12}) we see that the average of some combination of powers of $D_i$ remains constant (to
logarithmic  accuracy) if the corresponding $w_{\xi}$ is equal to zero. So, we are interested in
those sets $(m_1,\dots, m_d)$ that provide $w_{\xi}(m_1,\dots,m_d)=0$.

 Statistical isotropy and
incompressibility of the flow require $\d w_A / \d A_{ii} (0)=\frac 1{d} \langle {\rm tr} \bA
\rangle =0$. Since $w_A(0)=0$ and $w_A$ is concave, this means that for all non-zero arguments
$w_A$ is positive.

\begin{figure}[h]
\hspace*{-0.5cm} \includegraphics[width=8.5cm]{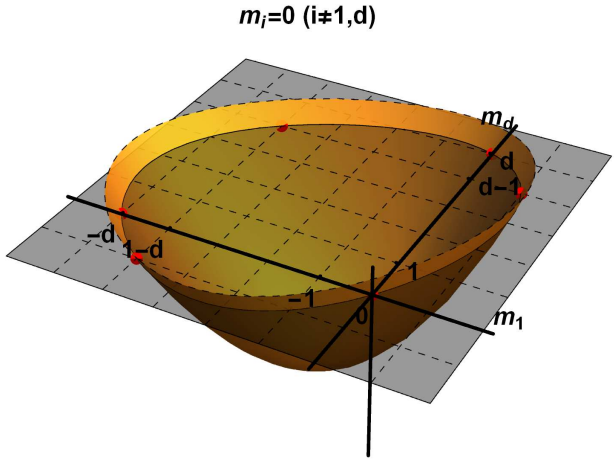}
\caption{Illustration of the surface $w_{\xi} (m_1,0,\dots,0,m_d)$  for $d\ge 3$ (yellow surface). The transection with the grey plane singles out the line $w_{\xi}=0$. For all the points $(m_1,m_d)$ in this line, $\langle D_1^{m_1} D_d^{m_d} \rangle $   conserve for given statistics of $\bA$. The red points belong to the line $w_{\xi}=0$ independently of the statistics; the corresponding  sets $\{ m_1^*, m_d^* \}$  produce universal integrals  of motion. }
\end{figure}

The function $w_{\xi}(m_1,\dots,m_d)$  can be obtained from $w_A$ in accordance with (\ref{main});
so, for all those points $(m_1^*,\dots,m_d^*)$ for which %
$ w_A(m_1^*+\eta_1,\dots,m_d^*+\eta_d) =w_A(\eta_1,\dots,\eta_d) \equiv w_A^* $ %
we have (Fig. 1)
\begin{equation} \label{wxi}
w_{\xi} (m_1^*, \dots,m_d^*) =0
\end{equation}
Each of these points corresponds to some stochastic integral of motion: actually,
$$
\left \langle D_1^{m_1^*}\dots D_d^{m_d^*} \right \rangle  = const
$$
In terms of the Cartan forms (\ref{norms}),  with account of (\ref{s-D}) this can be written as
$$
\left \langle s_1^{m_1^*-m_2^*} \dots  s_{d-1}^{m_{d-1}^*-m_d^*} s_d^{m_d^*} \right \rangle = const
$$
Thus, for any stationary Lagrangian statistics of velocity gradients, there exists a $d!-1$
-parametric family of stochastic invariants composed of $s_i$.

The solution of Eq.(\ref{wxi}) depends on the statistics; the powers $m_i^*$ for any flow are
determined by the specific form of the function $w_A$. However, it appears that all possible
surfaces $w_{\xi}(m_1^*,\dots, m_d^*)=0 $  corresponding to different statistics have several
points where they all intersect.  Indeed, since the process $\bA(t)$ is isotropic, all its diagonal
elements have identical statistical properties, and $w_A$ is symmetric with respect  to permutation
of its arguments: \\
if $\pi : 1\to \pi(1), \dots, d\to \pi(d)$ is a permutation  then 
\begin{equation} \label{permut}
 w_A (m_1,\dots, m_d) = w_A (m_{\pi(1)},\dots,m_{\pi(d)})
 \end{equation}
In particular, this is valid for the set $\{ \eta_1, \dots, \eta_d \}$. Hence, independently of the
details of statistics, the set $m_i^*= \eta_{\pi(i)} - \eta_i$ is the solution of $(\ref{wxi})$;
making use of (\ref{eta}),we find $m_i^* = i - \pi(i)$, and
\begin{equation} \label{with-sd}
\left \langle s_1 ^{\pi(2)-\pi(1)-1} s_2 ^{\pi(3)-\pi(2)-1}...s_{d-1} ^{\pi(d)-\pi(d-1)-1} s_d
^{d-\pi(d)} \right \rangle  = const
 \end{equation}
With account of incompressibility, $s_d=const$, and we arrive at (\ref{result}).

There are $d!$ permutations of the set (\ref{eta}); identical transform $\{\eta_i\} \to\{\eta_i\} $
corresponds to the trivial invariant, $\langle 1 \rangle =1$; for the rest $d!-1$ permutations we
get the same number of  integrals of motion.

\section{VI. Discussion}
So, in this paper we find $d!-1$ universal stochastic integrals of motion expressed in terms of
Cartano forms, or infinitesimal material lines and hyper-surfaces in a turbulent flow. The
universality means that the explicit forms of the invariants do not depend on the details of
velocity statistics. The only requirement is that all correlators of the velocity gradient tensor
along the trajectory of any particle are  isotropic and stationary, with finite correlation time,
and  independent of the choice of a particle. This requirement holds for incompressible isotropic
and homogenous flows with finite time and length correlation.

From isotropy of the flow it follows that one can take an average along an arbitrary generic line
or (hyper)surface instead of the ensemble average (the characteristic scale of the surface must be
much more than the correlation length). Thus, in (\ref{result}) it is possible to replace the average over an ensemble of liquid particles with an average taken over some material line or surface, or even
the whole space: in terms of the Lagrangian (frozen) coordinates~$\bx$,

\begin{equation} \label{disc1}
\int  s_1 ^{\pi(2)-\pi(1)-1} s_2 ^{\pi(3)-\pi(2)-1}...s_{d-1} ^{\pi(d)-\pi(d-1)-1}
d \bx = const
\end{equation}
(The integral can be taken over any subset of the coordinates $\{ x_1,\dots,x_d\}$.) Stochastic
invariance implies that the integral does not change exponentially as a function of time; it may
still have a power-law time dependence, which is a result of pre-exponential multipliers in $\bf
D$.

For the particular case of cyclic permutations we  get
$$
\left \langle s_k^{-d} \right \rangle = const
$$
According to (\ref{disc1}),
  these invariants  can be also written as integrals over a material element. Let a $k$-dimensional
material hypersurface $\{ x_1,\dots, x_k \}$  be marked by a passive scalar; let its initial
hypersurface density be uniform, $\sigma_k (0,\bx)=1$. Then, as time goes, the  hypersurface
density changes in accordance with the change of the hypersquare:  $\sigma_k (t,\bx) = 1/s_k (t,
\bx)$. We choose a fragment of the hypersurface; let its initial hypersquare be $S_0$. Then, making
use of (\ref{ldef}),(\ref{norms}) we pass from the integration over the Lagrangian coordinates to
the integration over the invariant measure (i.e., over the square of the hypersurface): $\int
dx_1\dots
dx_k = \int dS s_k^{-1}$. Now, 
one can write the integral of motion in the form
$$
\left \langle s_k^{-d} \right \rangle = \lim \limits _{S_0 \to \infty} \frac 1{S_0} \int d x_1\dots
dx_k \sigma_k ^d = \lim \limits _{S_0 \to \infty} \frac 1{S_0} \int d{S} \sigma_k ^{d+1}
$$
So, we arrive at (\ref{densities}).

This interpretation of the stochastic integrals allows a visualization that illustrates their
relation to the intermittency. Let a $k$-dimensional hypersurface be initially marked uniformly by
some scalar ('paint'). As time goes, the hypersurface undergoes deformations, stretches and bends.
The average density of the 'paint' decreases exponentially, inversely to the increase of  the
square. However there are always some rare and small regions where the density increases. The
balance between the small number of these regions and the very high density of 'paint' in them
results in the existence of the invariants (\ref{densities}). The higher-order statistical moments
grow, while the lower-order moments decrease as a function of time.

Returning back to Eq. (\ref{result}), we now present a recurrent  procedure to obtain the complete
set of these invariants in $d+1$-dimensional space from the set of the invariants in $d$
dimensions.

First, we note that, being written in the form (\ref{with-sd}), which includes the multiplier
$s_d$, the '$d$-dimensional' invariant is at the same time the invariant for $(d+1)$-dimensional
case; it corresponds to the permutations $\{ \pi(1),\dots, \pi(d),d+1\}$. We now write it in the
form
$$
\begin{array}{l}
\left \langle s_1^{\alpha_1}\dots s_d^{\alpha_{d}} s_{d+1}^{\alpha_{d+1}} \right \rangle \ , \\
\alpha_k = \pi (k+1) - \pi(k) -1 \ , \ \ 1 \le k \le d \ , \  \alpha_{d+1}=0
\\
\pi(d+1)=d+1
\end{array}
$$
Second, we make a cyclic permutation of the set $\{\pi(1),\dots, \pi(d+1)\}$ shifting it  by $i$:
$$
\tilde{\pi} (k) =  \left\{
\begin{aligned}
&\pi(k-i+d+1) \ ,&& \ k < i
 \\   &  \pi(k-i) \ ,&&\ k>i
 \end{aligned}
\right.
$$
In accordance with (\ref{result}) for $d+1$ dimensions, this new $\tilde{\pi} (k)$  corresponds to
$\langle s_1^{\beta_1} \dots s_d ^{\beta_d} \rangle $  where
$$
\begin{array}{l}
\beta_k = \tilde{\pi}({k+1}) - \tilde{\pi}(k) -1 = \left\{
\begin{array}{l}
\alpha_{k-i+d+1} \ , \ \ k<i \\ \alpha_{k-i} \ , \ \ i < k \le d
 \end{array}
\right.
\\
\beta_i = \pi(1)-(d+1)-1 = -\sum \limits_1^{d} \alpha_i - (d+1)
\end{array}
$$
This set of $\beta_k$ determines the new $(d+1)$-invariant. Since any permutation
 of $\{1,\dots,d+1 \}$
is a cyclic permutation of some $\{\pi(1),\dots, \pi(d),d+1\}$, this procedure allows us to find
all the $(d+1)$-invariants from known $d$-invariants.

If one wants to continue this recursion, one has to restore the form (\ref{with-sd}), i.e. to find
the power $\beta_{d+1}$ of $s_{d+1}$. To this purpose, it is helpful to use the (evident from
(\ref{with-sd})) property $\beta_1 + 2\beta_2 + \dots + (d+1) \beta_{d+1} =0$ for
$(d+1)$-dimensional invariants.

For example, in two dimensions we have one non-trivial permutation: (2,1), with corresponding
invariant $\langle s_1^{-2}s_2^1 \rangle$; in three dimensions we rewrite it as $\langle s_1^{-2}
s_2^{1} s_3^0\rangle $, corresponding permutation is 213. The further cyclic permutations lead to
(321,$i=1$) $\langle s_1^{-2} s_2^{-2} \rangle = \langle s_1^{-2} s_2^{-2}s_3^2 \rangle $  and
(132,$i=2$) $\langle s_1^{1} s_2^{-2} \rangle = \langle s_1^{1} s_2^{-2}s_3^1 \rangle $. The other
invariants come from the cyclic permutations of the ordered set (1,2,3). 
Figure 2 presents this procedure with a list of corresponding stochastic integrals of motion for the 3-dimensional case.

\begin{figure}[h]
\hspace*{-0.5cm} \includegraphics[width=8.5cm]{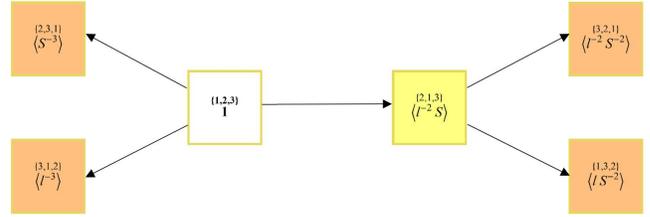}
\caption{Integrals of motion for incompressible flow in three dimensions and the illustration of the recurrent procedure. Each square would generate 3 more squares in the case of four dimensions.}
\end{figure}

The developed mechanism allows also to construct invariants composed of  two and more averages.
Actually, the permutation properties of $w_A$ are not restricted to the sets $\{m_i^*\}$. From
(\ref{main}) and (\ref{permut}) it follows that for any set $\{ m_i \}$, the set
$\{m'_i=m_{\pi(i)}+\eta_{\pi(i)}-\eta_i \}$ provides the same $w_{\xi}$,
 $w_{\xi}(m_1,\dots,m_d)=w_{\xi}(m'_1,\dots,m'_d)$, and the corresponding moments grow with the same rate.
  Hence, the ratio
 $\langle D_1^{m'_1}\dots D_d^{m'_d} \rangle / \langle D_1^{m_1}\dots D_d^{m_d} \rangle $ does
 neither increase nor decrease exponentially, and we get one more set of stochastic invariants:
 $$
\frac{\left \langle s_1^{m'_1-m'_2}\dots s_{d-1}^{m'_{d-1}-m'_d} \right \rangle }
 {\left \langle s_1^{m_1-m_2}\dots s_{d-1}^{m_{d-1}-m_d}  \right \rangle } = const \ , \
 m'_i  = m_{\pi(i)} + i - \pi(i)
 $$

Intermittent nature of a turbulent flow provides a wide range of invariants, some of them are
universal and independent of statistic properties of the flow. The formalism of generalized
Lyapunov exponents  appears to be a useful tool to find them.

\vspace{0.7cm}

We are grateful to Prof. A.V. Gurevich for his permanent interest to our work and constant support.

\end{document}